\documentclass[reprint,
amsmath,amssymb,
aip, jcp,
superscriptaddress,
longbibliography]{revtex4-1}

\usepackage[]{graphicx}
\usepackage{dcolumn}
\usepackage{bm}
\usepackage{hyperref}
\usepackage{color}
\usepackage{epstopdf}

\begin{document}
\title{The role of attraction in the phase diagrams and melting scenarios of generalized 2D Lennard-Jones systems}

\author{Elena N. Tsiok}
\affiliation{Institute of High Pressure Physics RAS, 108840
Kaluzhskoe shosse, 14, Troitsk, Moscow, Russia}
\author{Yuri D. Fomin}
\affiliation{Institute of High Pressure Physics RAS, 108840
Kaluzhskoe shosse, 14, Troitsk, Moscow, Russia}
\affiliation{Moscow Institute of Physics and Technology (National Research University), 9 Institutskiy Lane, Dolgoprudny, Moscow region, 141701, Russia}
\author{Evgenii A. Gaiduk}
\affiliation{Institute of High Pressure Physics RAS, 108840
Kaluzhskoe shosse, 14, Troitsk, Moscow, Russia}
\author{Elena E. Tareyeva}
\affiliation{Institute of High Pressure Physics RAS, 108840
Kaluzhskoe shosse, 14, Troitsk, Moscow, Russia}
\author{Valentin N. Ryzhov}
\email{ryzhov@hppi.troitsk.ru}
\affiliation{Institute of High Pressure Physics RAS, 108840
Kaluzhskoe shosse, 14, Troitsk, Moscow, Russia}
\author{Pavel A. Libet}
\affiliation{Institute of High Pressure Physics RAS, 108840
Kaluzhskoe shosse, 14, Troitsk, Moscow, Russia}
\affiliation{Bauman Moscow State Technical University, 2nd Baumanskaya street 5, 105005
Moscow, Russia}
\author{Nikita A. Dmitryuk}
\affiliation{Bauman Moscow State Technical University, 2nd Baumanskaya street 5, 105005
Moscow, Russia}
\author{Nikita P. Kryuchkov}
\affiliation{Bauman Moscow State Technical University, 2nd Baumanskaya street 5, 105005
Moscow, Russia}
\author{Stanislav O. Yurchenko}
\email{st.yurchenko@mail.ru}
\affiliation{Bauman Moscow State Technical University, 2nd Baumanskaya street 5, 105005
Moscow, Russia}

\date{\today}

\begin{abstract}
Monolayer and two-dimensional (2D) systems exhibit rich phase
behavior, compared with 3D systems, in particular, due to the
hexatic phase playing a central role in melting scenarios. The
attraction range is known to affect critical gas-liquid behavior
(liquid-liquid in protein and colloidal systems), but the effect
of attraction on melting in 2D systems remains unstudied
systematically. Here, we reveal how the attraction range affects
the phase diagrams and melting scenarios in a 2D system. Using
molecular dynamics simulations we considered the generalized
Lennard-Jones system with a fixed repulsion branch and different
power indices of attraction, from long-range dipolar to
short-range sticky-spheres-like. A drop in the attraction range
has been found to reduce the temperature of the gas-liquid
critical point, bringing it closer to the gas-liquid-solid triple
point. At high-temperatures, attraction does not affect the
melting scenario that proceeds through the cascade of
solid-hexatic (Berezinski-Kosterlitz-Thouless) and hexatic-liquid
(first-order) phase transitions. In the case of dipolar
attraction, we observed \emph{two triple points}, inherent in a 2D
system: hexatic-liquid-gas and crystal-hexatic-gas, the
temperature of the crystal-hexatic-gas triple point is
\emph{below} the hexatic-liquid-gas triple point. This observation
may have far-reaching consequences for future studies, since phase
diagrams determine possible routes of self-assembly in molecular,
protein, and colloidal systems, whereas the attraction range can
be adjusted with complex solvents and external electric or
magnetic fields. The results obtained may be widely used in
condensed matter, chemical physics, materials science, and soft
matter.
\end{abstract}


\maketitle

\section{Introduction}

Understanding phase transitions in 2D systems has prominent
importance in a number of areas, from photonics and electronics,
to novel materials and biotechnologies, since the knowledge on
phase behavior opens a way to design systems with desired
properties. Despite many studies, the fundamental questions here
are still related to the effect of particular interaction between
individual particles on their collective behavior. For classical
systems, one of the simplest models capable of reproducing their
behavior, including gas, liquid, and solid phases, is the
Lennard-Jones (LJ) system. The LJ model is widely used for
analysis of behavior in molecular, protein, polymer, emulsion, and
colloidal soft matter \cite{book.fernandez}. The generalized LJ
potential (or the LJn-m-potential, where indices $n$ and $m$ are
responsible for the algebraic branches of repulsion and
attraction, see below) is a suitable model for studies, aimed to
reveal the effects of repulsion and attraction on liquids and
solids, and the phase transitions between them.

Today, it has been established that 2D melting scenarios depend on
\emph{repulsion softness}, providing the following microscopic
scenarios of 2D melting \cite{10.3367/ufne.2017.06.038161,
10.3367/ufne.2018.04.038417}: (i) the
Berezinskii-Kosterlitz-Thouless-Halperin-Nelson-Young (BKTHNY)
theory, according to which melting occurs via two continuous
transitions with an intermediary hexatic phase with
quasi-long-range orientational order and short-range translational
order \cite{Berezinsky:1970fr, 10.1088/0022-3719/6/7/010,
10.1103/physrevlett.41.121, 10.1103/physrevb.19.2457,
10.1103/physrevb.19.1855}; (ii) melting through a first order
phase transition \cite{10.1103/physrevb.28.178, ryzhov1991}; (iii)
two-step melting, including a continuous
(Berezinskii-Kosterlitz-Thouless, BKT) crystal-hexatic phase
transition and a first-order phase transition between the hexatic
phase and isotropic liquid \cite{10.1103/physrevlett.107.155704,
10.1103/physreve.87.042134, 10.1039/c4sm00125g,
10.1103/physrevlett.114.035702, 10.1103/physrevlett.118.158001}.
The second and third scenarios are inherent in hard-sphere-like
systems, whereas the first one was observed for soft repulsion
between particles \cite{WOS:000344847600066, WOS:000338123700024,
WOS:000179502800054}. It has been established unambiguously that
the softness of repulsion affects melting scenarios,
thermodynamics and excitation spectra in monolayer systems
\cite{WOS:000360886700006, WOS:000173912900008,
10.3367/ufne.0182.201211a.1137, 10.1063/1.4979325,
10.1103/physreve.97.022616, 10.1103/physrevlett.121.075003,
10.1038/s41598-019-46979-y, 10.1021/acs.jpclett.9b01468,
10.1021/acs.jpclett.9b03568, 10.1103/physrevlett.125.125501,
10.1103/physreve.103.052117, WOS:000444571600047,
WOS:000530499300009, WOS:000665626600004}. However, to the best of
our knowledge, the role of \emph{attraction} in the melting
scenario of monolayer systems remains unstudied systematically.

LJ interactions were among the first systems studied to understand
the role of attraction in melting. However, a lot of reported
results on the critical point and melting scenario for 2D LJ
crystals still remain questionable. For instance, to provide
critical temperature depending on the truncation radius, numerical
simulations of a vapor-liquid curve performed in the Gibbs
ensemble were reported in Ref.~\cite{10.1063/1.460477}. For
critical temperature and density, the authors obtained $T_c =
0.515 \pm 0.002$ and $\rho_c = 0.355 \pm 0.003$, respectively, for
the full potential; and $T_c = 0.459 \pm 0.001$ and $\rho_c = 0.35
\pm 0.01$ for the truncated and shifted potential at $2.5\sigma$.
Contradictory melting scenarios of the triangular crystal were
reported in the early works \cite{10.1103/physrevlett.42.1632,
10.1063/1.436526, 10.1103/physrevlett.44.463, 10.1063/1.441901,
10.1103/physrevlett.52.449, 10.1103/physrevb.30.2755}, including
two continuous transitions with an intermediate hexatic phase
according to the BKTHNY theory \cite{10.1103/physrevlett.42.1632}
and a first-order transition \cite{10.1063/1.436526,
10.1103/physrevlett.44.463, 10.1063/1.441901,
10.1103/physrevlett.52.449, 10.1103/physrevb.30.2755}.

Thanks to growth of computing capabilities, simulations of large
systems ($\gtrsim 10^5$ particles) have recently provided new
results on 2D melting of LJ crystals and related systems.
Simulations of the systems followed by analysis of their equation
of state and long-range asymptotics of the translational
correlation function (that accurately provides the stability limit
of the crystal) allowed identifying the melting scenarios
unambiguously. For instance, a change in the melting scenario was
reported in Ref.~\cite{10.1103/physreve.99.022145}, where the
authors studied 2D systems of particles interacting via
generalized LJ potential with different repulsive branches
($\propto 1/r^{12}$ and $\propto 1/r^{64}$). The scenario was
found to occur through first-order phase transitions at low
temperatures and via two continuous BKT transitions (according to
the BKTHNY theory) at high. An LJ system at high temperatures is
typically assumed to be close to soft repulsive disks $1/r^{12}$,
but such extrapolation to a melting scenario contradicts the
results of Ref.~\cite{10.1103/physrevlett.114.035702}, wherein the
soft disks $1/r^n$ with $n>6$ were shown to melt according to the
third melting scenario. The Mayer-Wood loop, inherent in
first-order transition, is assumed to disappear at high
temperatures with an increase in the system size. However, the
explanation of the effect by finite-size scaling seems
unconvincing: With an increase in the system size, the loop should
flatten and ultimately approach the
plateau~\cite{10.1103/physreve.87.042134,
10.1103/physreve.59.2659}.

A first-order hexatic-liquid transition and a continuous
crystal-hexatic BKT transition at high temperatures and one
first-order crystal-liquid transition at low temperatures were
identified in Ref.~\cite{10.1103/physreve.102.062101} for 2D LJ
particles. The results of numerical simulation of a 2D LJ system
and attractive polygons (squares, pentagons and hexagons) of the
same authors in Ref.~\cite{10.1103/physrevlett.124.218002}
revealed the role of attraction in a melting scenario. Thus, at
low temperatures where the role of attraction is dominant all
systems melt via first-order transition due to suppression of the
hexatic phase. At high temperatures LJ disks melt in accordance
with the third scenario, the same as soft disks
\cite{10.1103/physrevlett.114.035702, WOS:000415912900074,
10.1080/00268976.2019.1607917}, whereas hexagons and squares do
according to the BKTHNY theory with participation of the hexatic
and tetratic phases, respectively. The melting scenario of
pentagons does not change with an increase in temperature and is a
first-order transition.

LJ crystals compared with the Morse system in
Ref.~\cite{10.1103/physrevb.103.094107} were shown to melt via the
third scenario at low temperatures and two continuous BKT
transitions at high. This agrees with
Ref.~\cite{10.1103/physreve.99.022145}, but contradicts
Refs.~\cite{10.1103/physrevlett.114.035702,
10.1103/physreve.102.062101}. The BKTHNY scenario at high
temperatures was questioned because of seeming disappearance of
the Mayer-Wood loop (an analog of the Van-der-Waals loop in the 3D
case). For soft Morse interactions, the third melting scenario was
observed for all temperatures considered in
Ref.~\cite{10.1103/physrevb.103.094107}, whereas the authors
expected to observe the BKTHNY scenario at higher temperatures.
However, at some parameters of potential softness, two continuous
transitions were observed already at low temperatures in the
presence of long-range attraction.

The role of attraction can be tested experimentally in colloidal
systems, known for a long time as model systems demonstrating a
wide range of ``molecular-like'' phenomena \cite{book.fernandez,
book.ivlev, 10.1016/0370-1573(94)90017-5,
10.1038/natrevmats.2015.11, 10.1039/c9sm01953g}, in particular,
crystallization and melting \cite{10.1126/science.1112399,
10.1039/c2sm26473k, 10.1103/physrevlett.82.2721,
10.1103/physrevlett.85.3656, 10.1103/physrevlett.118.088003,
10.1039/c2sm27654b, 10.1126/science.1224763, 10.1063/1.5116176,
10.1038/s41598-021-97124-7}, solid-solid phase transitions
\cite{10.1063/1.2189850, 10.1038/nature01328, 10.1038/nmat4083},
condensation and critical phenomena \cite{10.1038/nphys679,
10.1039/c2sm27119b, 10.1103/physrevx.9.031032} caused by
relatively \emph{short-range} attractive depletion
forces\cite{book.Lekkerkerker}. These collective phenomena can be
visualized \emph{in real-time} with spatial resolution of
individual particles. \emph{Long-range} dipolar attraction
$\propto 1/r^3$ in colloidal systems can be induced and controlled
\emph{in situ} with in-plane rotating magnetic
\cite{10.1088/0034-4885/76/12/126601, 10.1039/c3sm50306b,
10.1039/c3sm27620a, 10.1039/c6sm02131j,
10.1103/physrevmaterials.2.025602} or electric
\cite{10.1088/1367-2630/8/11/267, 10.1063/1.3115641,
10.1021/la2014804, 10.1021/la500178b, 10.1039/c1sm06414b,
10.1038/s41598-017-14001-y, 10.1063/1.5131255,
10.1016/j.jcis.2021.09.116} fields. Using conically-rotating
magnetic or electric fields with magic angles,
\emph{Van-der-Waals-like attraction} can be created $\propto
1/r^6$ with ``magic'' fields \cite{10.1021/la500896e,
10.1103/physrevlett.103.228301, 10.1103/physreve.95.052607,
10.1039/c8sm01538d}. Recently, tunable interactions have been
achieved by using spatial hodographs of an external electric or
magnetic field \cite{10.1039/d0sm01046d}, by engineering the
internal structure\cite{10.1063/5.0055566} and
geometry\cite{10.1063/5.0060705} of colloidal particles.

In the present paper, we studied the role of the attraction range
in the phase behavior and melting scenarios of a 2D system. Using
MD simulations, we considered a generalized Lennard-Jones system
with a fixed repulsion branch and different attraction, from
long-range dipolar attraction to short-range sticky-spheres-like.
A decrease in the attraction range is found to suppress the
temperature of the gas-liquid (liquid-liquid in the case of
colloid and protein systems) critical point. We found that
attraction does not affect melting at high temperatures, occurring
through the cascade of solid-hexatic (BKT) and hexatic-liquid
(first-order) phase transitions. Surprisingly, at moderate
temperatures we observed \emph{two triple points},
hexatic-liquid-gas and crystal-hexatic-gas. The
crystal-hexatic-gas triple point for the case of isotropic dipolar
attraction has been discovered to be \emph{below} the
hexatic-liquid-gas triple point.

\section{System and methods}

\subsection{MD details}
We studied a system of particles interacting via the generalized
Lennard-Jones potential (LJn-m):
\begin{equation}
\label{NMP-eq1}
U_{nm}(r)=\frac{\epsilon}{n-m}\left[m\left(\frac{\sigma}{r}\right)^n- n\left(\frac{\sigma}{r}\right)^m\right],
\end{equation}
where $n$ and $m$ are the indices of repulsive and attractive
branches, respectively, and $\sigma$ and $\epsilon$ are the
characteristic length of interaction and the depth of the
potential well. The potential \eqref{NMP-eq1} has minimum
$-\epsilon$ at $r/\sigma=1$. In what follows, we normalized the
distances and energies to $\sigma$ and $\epsilon$, respectively,
and considered the particles of equal mass $m=1$.

\begin{figure}[!t]
    \includegraphics[width=80mm]{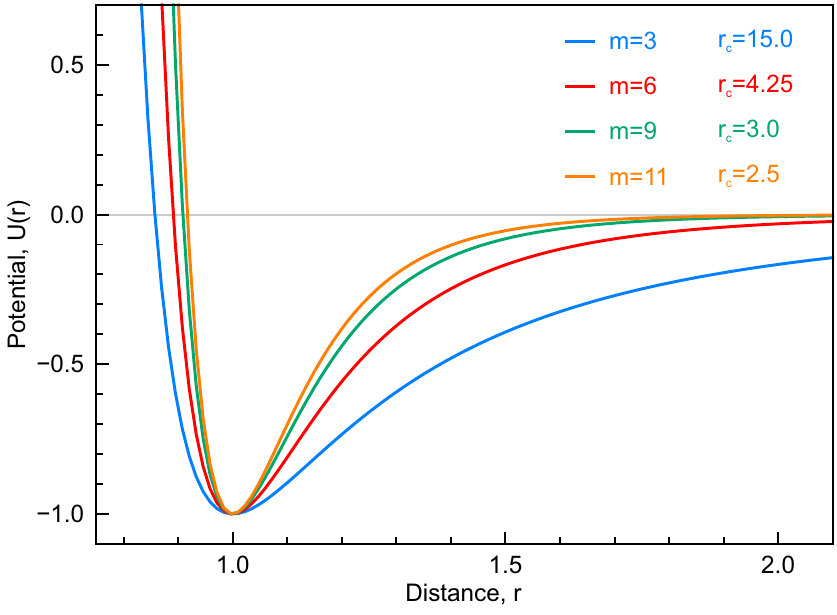}
    \caption{\textbf{Generalised LJ potentials with different attraction indices:}
    The solid lines correspond to fixed repulsion with $n=12$ and different indices $m=3$, $6$, $9$, and $11$, illustrating the transition from long- to short-range attractive interactions.
    The values of cutoff radii $r_c$ (used for calculations of the equation of state) decrease as shown for each $m$. }
    \label{NMP-Figure1}
\end{figure}

We considered cases with fixed repulsion branch $n=12$ and
different attraction, from long-range isotropic dipolar attraction
with $m=3$ to short-range almost sticky-spheres-like attraction
with $m=11$. Examples of the potentials with $m=3$, $6$, $9$, and
$11$ are given in Fig.~\ref{NMP-Figure1}. A case similar to usual
LJ interaction corresponds to $n=12$ and $m=6$.

\begin{table}[!b]
    \caption{\textbf{Simulation parameters for potentials LJ12-m used for calculations in the phase identification method, where $a = \rho^{-1/2}$}}
    \centering{
    \begin{tabular}{c|c|c|c|c|c}
        LJn-m   & $\rho$ & $r_c$  & $T_{start}$ & $T_{stop}$ & $T_{step}$\\ \hline
        LJ12-3  & 0.3    & 7.5$a$ & 0.7         & 3.05       & 0.06      \\
        LJ12-4  & 0.4    & 7.5$a$ & 0.1         & 2.0        & 0.03      \\
        LJ12-5  & 0.4    & 7.5$a$ & 0.1         & 1.5        & 0.02      \\
        LJ12-6  & 0.4    & 7.5$a$ & 0.1         & 1.5        & 0.02      \\
        LJ12-7  & 0.4    & 7.5$a$ & 0.454       & 0.503      & 0.003     \\
        LJ12-8  & 0.4    & 7.5$a$ & 0.2         & 0.4        & 0.003     \\
        LJ12-9  & 0.4    & 7.5$a$ & 0.14        & 0.3        & 0.002     \\
        LJ12-10 & 0.4    & 5.0$a$ & 0.09        & 0.151      & 0.001     \\
        LJ12-11 & 0.5    & 5.0$a$ & 0.025       & 0.076      & 0.001     \\
    \end{tabular}
    \label{NMP-Table1}
    }
\end{table}

The molecular dynamics (MD) simulations were performed for
$N=2\times10^4$ particles using LAMMPS in the canonical ($NVT$)
ensemble in a wide range of densities and temperatures, from
$T=0.3$ to $T=50.0$, using $10^8$ steps with time step $dt=
10^{-3}$ (in dimensionless units of time normalized to $\sqrt {m
\sigma^2 / \epsilon}$). The value of potential energy within
cutoff radius $r_c$ was $|U(r_c)| < 5\times 10^{-4}$: it was
$U(r_c) = -4\times10^{-4}$ for $m=3$, $U(r_c) = -3.4\times
10^{-4}$ for $m=6$, $U(r_c) = -2\times 10^{-4}$ for $m=9$, $U(r_c)
= -3.2\times 10^{-4}$ for $m=11$. To determine accurately the
boundaries of two-phase areas, crystal melting was examined (at
some points) for large systems with $N=256^2$ and $512^2$
particles. The phase diagram was obtained using the dependence of
pressure on density (the equation of state) along the isotherms,
analysis of the radial distribution functions, orientational and
translational order parameters, and the corresponding correlation
functions.

\subsection{The method of phase identification}
The condensate-gas binodals were obtained using the method of
phase identification, proposed in
Ref.~\cite{10.1021/acs.jpcc.7b09317}. The method allows obtaining
a binodal in the coordinates density-temperature using analysis of
Voronoi cells in the system and, in particular, in terms of the
following order parameter \cite{10.1021/acs.jpcc.7b09317,
10.1038/s41598-021-97124-7}:

\begin{equation}
\label{NMP-eq3}
\begin{split}
&\lambda_{i} = \frac{1}{N_{ni}+1}\left(\sigma_{i}+\sum_{j=1}^{N_{ni}}{\sigma_{j}}\right),\\
&\sigma_{i} =\frac{1}{a_i N_{ni}}\sqrt{\sum_{j<k}^{N_{ni}}{(r_{ij}-r_{ik})^2}/2}, \quad r_{ij}=|\mathbf{r}_i-\mathbf{r}_j|,
\end{split}
\end{equation}
where $\mathbf{r}_i$ is the radius-vector of the $i$-th particle,
$N_{ni}$ is the number of its neighboring cells,
$a_i=\sqrt{S_i/\pi}$, $S$ is the area of a corresponding Voronoi
cell. The scalar $\lambda^2$-field has small values for condensed
(liquid and solid) phases, and becomes large at the interface
gas-condensate and in gaseous state. Then, analysis of
corresponding statistics of Voronoi cells allows obtaining
densities $n_{c}$ and $n_{g}$ of condensate and gas, respectively
(see Ref.~\cite{10.1021/acs.jpcc.7b09317} for details).

\begin{figure*}[!t]
    \includegraphics[width=175mm]{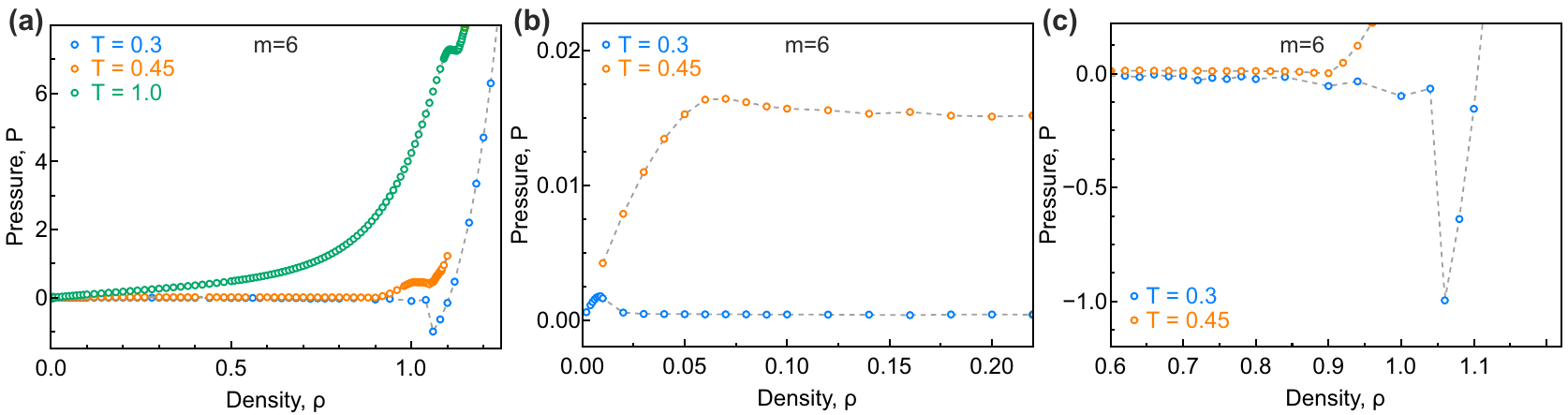}
    \caption{\textbf{The equation of state for the 2D LJ system:} (a) The isotherms of the system with $m=6$ below ($T=0.3$) and above ($T=0.45$) the triple point and above the critical point at $T=1.0$. (b) Isotherm $T=0.45$ gas-liquid and isotherm $T=0.3$ gas-crystal at low densities. (c) Isotherm $T=0.45$ gas-liquid and isotherm $T=0.3$ gas-crystal at high densities.}
    \label{NMP-Figure2}
\end{figure*}

To apply the method, additional simulations of a small system with
$N = 3.6 \times 10^3$ particles under the same $NVT$ conditions
(as we explained above), from $T_{start}$ to $T_{stop}$, with step
$T_{step}$ were conducted where each step included $6 \times 10^5$
time-steps with dimensionless time-step $\Delta t = 5
\times10^{-4}$ (see the parameters in Table~\ref{NMP-Table1}). At
each temperature, we performed phase identification using $10^2$
simulation frames.

\begin{table}[!b]
    \caption{\textbf{The temperatures and densities in the critical and triple points obtained using the phase identification method}}
    \centering{
    \begin{tabular}{c|c|c|c|c}
        LJn-m & $T_{CP}$  & $\rho_{CP}$ & $T_{TP}$ & $\rho_{TP}$\\ \hline
        LJ12-3 & 1.442 & 0.366 & 0.577 & 1.095 \\
        LJ12-4 & 0.748 & 0.453 & 0.455 & 1.042 \\
        LJ12-5 & 0.606 & 0.466 & 0.408 & 1.017 \\
        LJ12-6 & 0.510 & 0.488 & 0.390 & 1.012 \\
        LJ12-7 & 0.480 & 0.486 & 0.399 & 1.015 \\
        LJ12-8 & 0.436 & 0.509 & 0.390 & 1.016 \\
        LJ12-9 & 0.429 & 0.516 & 0.393 & 1.029 \\
        LJ12-10 & 0.410 & 0.544 & 0.386 & 0.996 \\
        LJ12-11 & 0.4053 & 0.550 & 0.3844 & 1.0478
    \end{tabular}}
    \label{NMP-Table2}
\end{table}

This method of phase identification becomes unsuitable near the
critical point, where the condensed and gaseous phases cannot be
distinguished. However, one can obtain the values of critical
density $n_{\mathrm{CP}}$ and temperature $T_{\mathrm{CP}}$ using
the following approximation of the binodal (the results are
provided in Table~\ref{NMP-Table2}):
\begin{equation}
\label{NMP-eq4}
n_{c}-n_{g} \simeq A \tau^{\beta}, \quad n_{c}+n_{g} \simeq a \tau+2 n_{\mathrm{CP}},
\end{equation}
where $\tau=T_{\mathrm{CP}}-T-$ is reduced temperature, $\beta$ is
a critical index, $A$ and $a$ are the free parameters of the
model. Critical index $\beta$ is related to the universality class
of the system and is determined by the range of
attraction\cite{10.1103/physrevlett.89.025703}: In 2D systems with
dipolar isotropic attraction, $\beta = 1/2$ (mean-field critical
behavior), whereas in the case of shorter-range attraction, at
$m>3$, $\beta = 1/8$ corresponds to 2D Ising
behavior\cite{10.1103/physrevlett.89.025703}.

\subsection{Order parameters}

For analysis of the phase transition scenarios, we used
orientational and translational correlation functions. The
orientational correlation function (OCF) of the global
orientational order parameter is calculated as:
\begin{equation}
\label{NMP-eq5}
G_6(r)=\frac{\left<\Psi_6(\mathbf{r})\Psi_6^*(0)\right>}{g(r)}, \quad
\Psi_6=\frac{1}{N}\left<\left|\sum_i \psi_6(\mathbf{r}_i)\right|\right>,
\end{equation}
where the averaging for $G_6$ is performed over all particles in
the system, $\psi_6 = 1/n(i) \sum_j{e^{6i\theta_{ij}}}$,
$\theta_{ij}$ is the angle of the vector between particles $i$ and
$j$ with respect to the reference axis, and the sum over $j$ is
counting the $n(i)$ nearest-neighbors of $j$, obtained from the
Voronoi construction,
$g(r)=\langle\delta(\mathbf{r}_i)\delta(\mathbf{r}_j)\rangle$ is
an isotropic pair distribution function (here, $\mathbf{r}_i$ is
the position vector of particle $i$, and
$r=|\mathbf{r}_i-\mathbf{r}_j|$). In the hexatic phase, $G_6(r)$
behaves at large distances like $G_6(r)\propto r^{-\eta_6}$ with
$\eta_6 \leq 1/4$ \cite{10.1103/physrevlett.41.121,
10.1103/physrevb.19.2457}.

The translational correlation function (TCF) is
\begin{equation}
\label{NMP-eq7}
G_t(r)=\frac{\langle \exp(i \mathbf{a}(\mathbf{r}_i-\mathbf{r}_j))\rangle}{g(r)},
\end{equation}
where $\mathbf{a}$ is the reciprocal-lattice vector of the first
shell of the crystal lattice. In the solid phase, $G_t(r)$ behaves
at large distances like $G_t(r)\propto r^{-\eta_T}$ with $\eta_T
\leq {1}/{3}$ \cite{10.1103/physrevlett.41.121,
10.1103/physrevb.19.2457}. In the hexatic phase and isotropic
liquid, $G_t$ decays exponentially.

\begin{figure}[!b]
    \includegraphics[width=80mm]{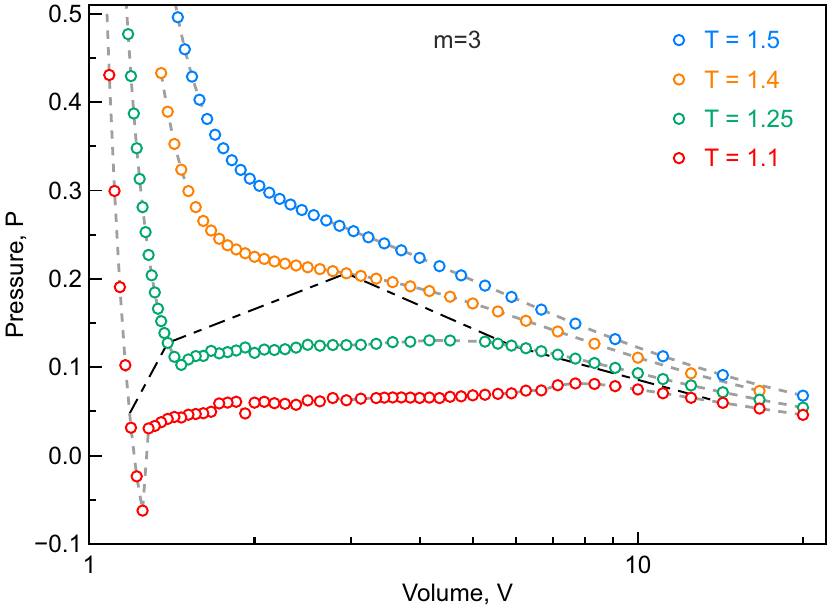}
    \caption{\textbf{The isotherms of LJ12-3 system in $P-V$ coordinates:}
    The gas-liquid transition boundary is obtained using Maxwell's construction at different temperatures and is shown with the black dashdotted line.}
    \label{NMP-Figure3}
\end{figure}

\section{Results and discussion}

\subsection{The equations of state and phase diagrams}

The equations of state (isotherms) for all considered systems
behave in a similar manner and have peculiarities that, depending
on the temperature and density, attributed to gas-liquid
transition or crystal melting, respectively. This is shown in
Fig.~\ref{NMP-Figure2}(a): Here, at isotherm $T=0.45$, one can see
a wide loop peculiar to gas-liquid transition and a narrow loop
related to crystal melting, whereas only one loop related to
crystal melting is seen above the gas-liquid critical point at
$T=1.0$. Below the triple point, at $T=0.3$, there is only one
wide loop corresponding to gas-crystal transition.

\begin{figure*}[!t]
    \includegraphics[width=160mm]{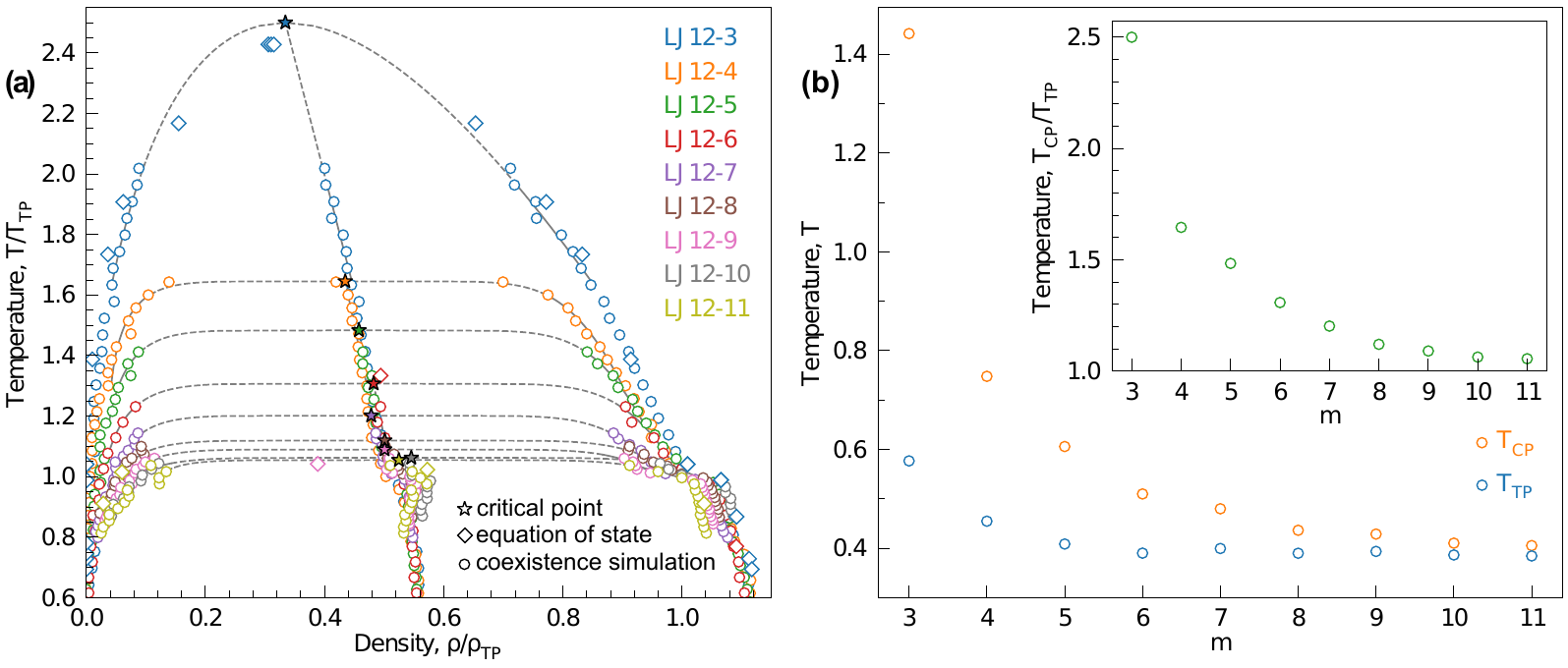}
    \caption{\textbf{The effect of the attraction range on the liquid-gas coexistence area on the phase diagram:}
    (a) The binodals condensate-gas for different LJ12-m potentials; the circles are the points of the binodal and median (obtained using the phase identification method), the diamonds are points obtained from the equation of state, the grey lines are the approximations of the binodal with Eq.~\eqref{NMP-eq4}, the stars denote the critical points from Table~\ref{NMP-Table2}.
    (b) The dependencies of triple and critical temperatures on attraction index $m$ for LJ12-m interaction, the ratio $T_{CP}/T_{TP}$ is shown in the inset.}
    \label{NMP-Figure4}
\end{figure*}

The binodals gas-liquid were calculated using Maxwell's
construction on the isotherms in the $P-V$ coordinates (an example
is shown in Fig. \ref{NMP-Figure3} for the system with $m=3$), as
well as with the phase identification method. The results for the
binodal condensate-gas obtained from the phase identification
method and the equation of state are provided in Fig.
\ref{NMP-Figure4}(a). Here, the colored circles denote the
densities of gas, condensate, and their average for each potential
we considered. The grey solid lines show the regions of the data
wherein we used the approximation \eqref{NMP-eq4} to obtain data
about the critical point. The grey dashed lines show the
extrapolation of the phase diagram to the critical points
gas-liquid, depicted with colored stars (the corresponding
densities and temperatures are given in Table~\ref{NMP-Table2}).

We observed that the drop in the attraction range reduces the
critical temperature, as well as the ratio between temperatures of
the critical and triple points, as shown in
Fig.~\ref{NMP-Figure4}(b) and the corresponding inset (based on
the data in Table~\ref{NMP-Table2}). With an increase in $m$
(short-range attraction), the two-phase area becomes narrower
towards lower densities and the ratio between the critical and
triple temperatures becomes closer to unity, as shown in
Fig.~\ref{NMP-Figure4}(b). For LJ interaction ($m=6$), the
critical temperature we obtained is $T_c = (0.51\ldots0.52)$
(depending on the method of evaluation), which agrees well with
the previous results $T_c = 0.515 \pm 0.002$ for the LJ potential
reported in Ref.~\cite{10.1063/1.460477}.

At high temperatures, the effect of attraction should vanish, and
only the repulsive branch of potential \eqref{NMP-eq1} should play
a more significant role. Indeed, analyzing the melting scenarios
of triangular crystals at high temperatures (and high densities),
we observed that LJ12-m systems demonstrated the same melting
scenarios as soft disks repulsive potential $1/r^{12}$ (the third
melting scenario~\cite{10.1103/physrevlett.114.035702,
10.1103/physreve.102.062101, 10.1080/00268976.2019.1607917}).

\begin{figure*}[!t]
    \includegraphics[width=160mm]{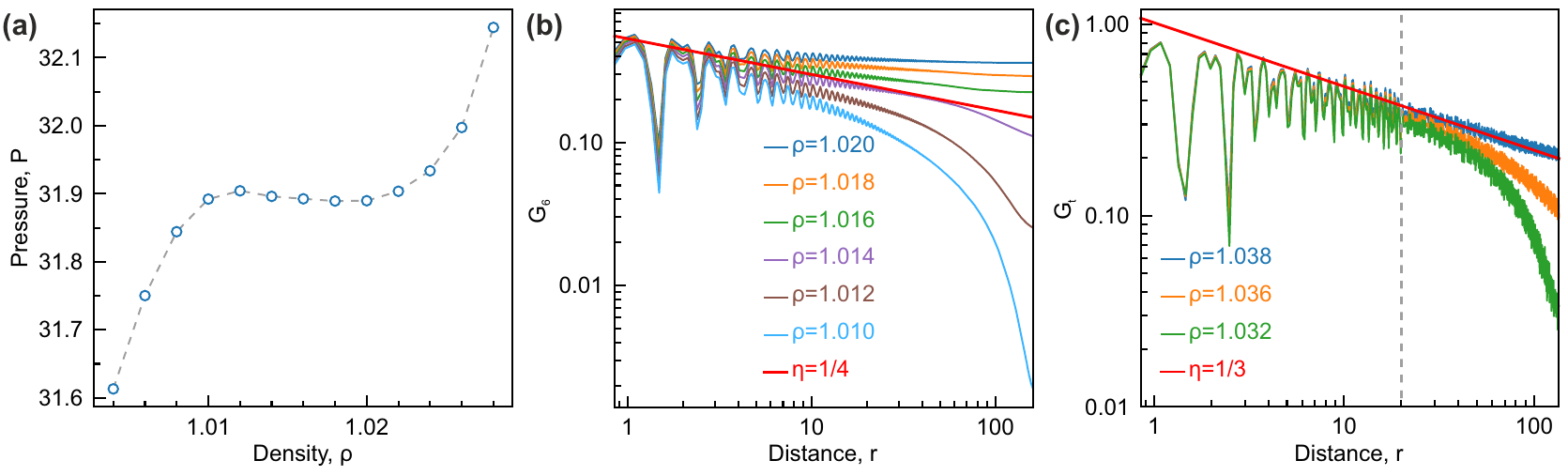}
    \caption{\textbf{Evolution of correlation functions on an isotherm:}
    (a) The isotherm $T=3.0$ for LJ system consisting of $N=512^2$, obtained by averaging over 20 independent replicas.
    (b) and (c) The OCF and TCF for the same system at different densities.}
    \label{NMP-Figure5}
\end{figure*}

\begin{figure}[!t]
    \includegraphics[width=80mm]{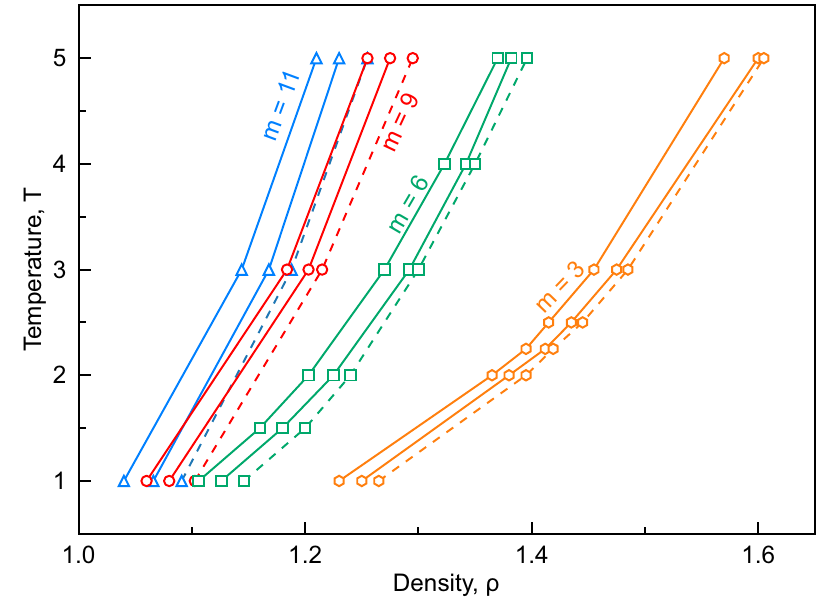}
    \caption{\textbf{The third melting scenario of the triangular crystals at high temperatures and densities:}
    The solid lines are the boundaries of the Mayer-Wood loop corresponding to the two-phase hexatic-liquid area.
    The dashed lines are the boundaries of BKT transition from the crystal to hexatic phase.
    The results are provided for the cases $m=3$, $6$, $9$, and $11$, as labeled in the figure.}
    \label{NMP-Figure6}
\end{figure}

The Mayer-Wood loop was observed in all considered systems at high
temperatures, proving first-order transition. However, to
determine an appropriate melting scenario, we compared the
stability limits of the crystal and the hexatic phases with the
boundaries of the Mayer-Wood loop. The stability limits of the
hexatic and crystal phases were derived from asymptotic behavior
of the correlation functions of the orientational and
translational order parameters, respectively. An example of
evolution of the correlation functions on the isotherm of the LJ
system at $T=3.0$ is shown in Fig.~\ref{NMP-Figure5}.

In Fig.~\ref{NMP-Figure5}(a), the equation of state is shown at
$T=3.0$ (obtained by averaging over 20 independent replicas) for
the LJ system consisting of $N=512^2$ particles. One can see the
clearly defined Mayer-Wood loop, inherent in first-order
transitions. Analyzing the correlation functions in
Figs.~\ref{NMP-Figure5}(b,c), we conclude that the loop
corresponds to the two-phase hexatic-liquid area, whereas the
melting scenario is the third, with a continuous BKT transition
from crystal to hexatic and a first-order transition from hexatic
to liquid. Here, the hexatic phase becomes unstable at
$\rho_{hex-liq} = 1.014$ (Fig.~\ref{NMP-Figure5}(b)), inside the
loop, whereas the crystal becomes unstable at $\rho_{sol-hex} =
1.037$ (Fig.~\ref{NMP-Figure5}(c)) getting far out of the loop.
The obtained results agree well with the melting scenario for
repulsive disks $1/r^{12}$ (reported in
Refs.~\cite{10.1103/physrevlett.114.035702,
10.1103/physreve.102.062101, 10.1080/00268976.2019.1607917}), but
do not exhibit two continuous BKT transitions in the LJ system at
high temperatures \cite{10.1103/physreve.87.042134,
10.1039/d0sm01484b}. The obtained results agree well with the
melting scenario for repulsive disks $1/r^{12}$ (reported in
Refs.~\cite{10.1103/physrevlett.114.035702,
10.1103/physreve.102.062101, 10.1080/00268976.2019.1607917}), but
do not exhibit two continuous BKT transitions in the LJ system at
high temperatures \cite{10.1103/physreve.87.042134,
10.1039/d0sm01484b}.

All triangular crystals, we studied, were found to melt at high
temperatures via the third scenario, irrespective to the
attraction index $m$. The melting lines in this region of the
phase diagram are shifted to low densities with an increase in $m$
and behave similarly to a system of soft disks $1/r^{12}$, as
shown in Fig.~\ref{NMP-Figure6}.

\begin{figure}[!t]
    \includegraphics[width=80mm]{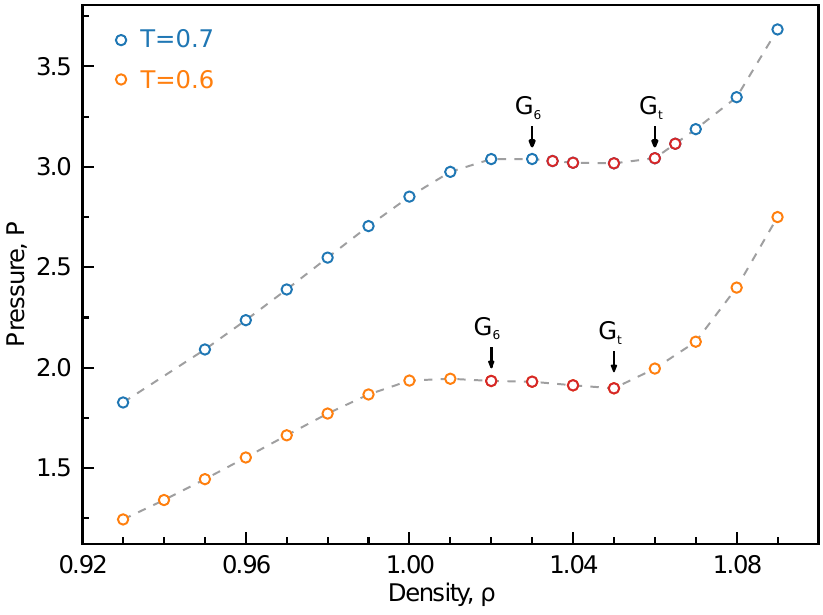}
    \caption{\textbf{The change in the melting scenario for the case $m=9$:}
    The melting scenario changes from a first-order transition ($T=0.6$) to the third scenario ($T=0.7$).
    The arrows show the stability limits of the hexatic phase, $G_6$ are obtained using the OCF, and the stability limits of the crystalline phase, $G_t$ are obtained with TCF, which become caught by the loop with a decrease in temperature.}
    \label{NMP-Figure7}
\end{figure}

However, with a decrease in melting temperature, the balance of
the effects provided by the repulsive and attractive branches is
changed. As a result, we observed that, at low temperatures, the
melting scenario \emph{changes} from the third one to a
first-order transition (without inclusion of the hexatic phase) in
the systems with $m=6$, $9$, and $11$. The change in the melting
scenario is shown for two isotherms of the LJ12-9 system in Fig.
\ref{NMP-Figure7}.

The temperature of the observed change in the melting scenario
decreases with $m$ (to more soft and long-range attraction). The
attraction range affects essentially the region of gas-liquid
transition, as well as temperatures $T_c$ and $T_m$, at which the
hexatic phase appears.

\subsection{The results for isotropic dipolar attraction}

\begin{figure}[!t]
    \includegraphics[width=80mm]{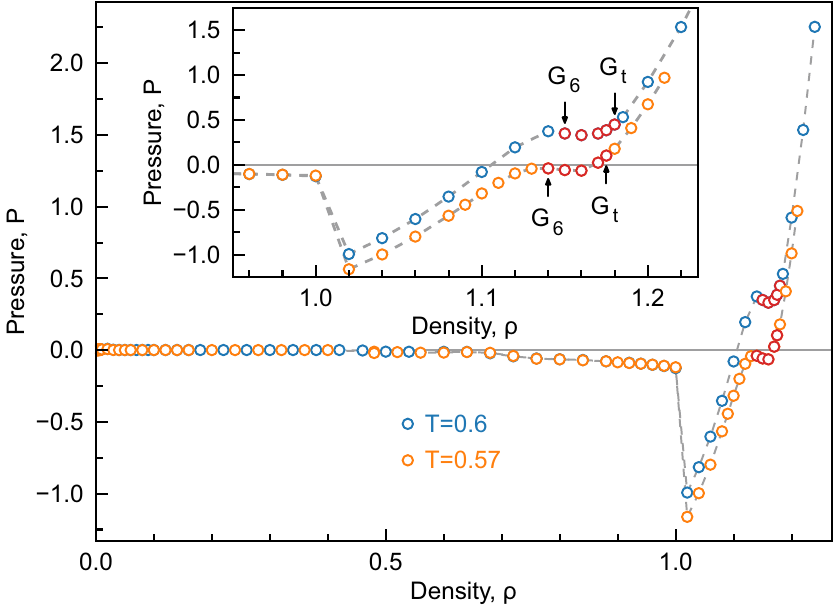}
    \caption{\textbf{The cascades of phase transitions:} Isotherm $T = 0.6 > T_t$ illustrates the cascade of a first-order gas-liquid transition, first-order liquid-hexatic transition, and a continuous BKT crystal-hexatic transition.
    At $T = 0.57 < T_t$, only a first-order gas-hexatic transition and a continuous BKT crystal-hexatic transition are seen.
    Detailed behavior of the isotherms at high densities are shown in the inset.
    At negative pressures and $T=0.57 < T_t$, one can see a loop related to the first-order transition gas-hexatic (in the metastable area), while a continuous BKT crystal-hexatic transition still occurs at positive pressure.
    The arrows show the limits of stability of the hexatic phase $G_6$ and of the crystal $G_t$, obtained using OCF and TCF analysis, respectively.}
    \label{NMP-Figure8}
\end{figure}

In the case of LJ12-3 potential, the crystal was found to melt at
low temperatures according to the third scenario. However, the
hexatic phase was discovered to be preserved up to triple point
temperature $T_t$, where it coexists with both liquid and gaseous
phases, as given in Fig.~\ref{NMP-Figure8}.

\begin{figure}[!t]
    \includegraphics[width=80mm]{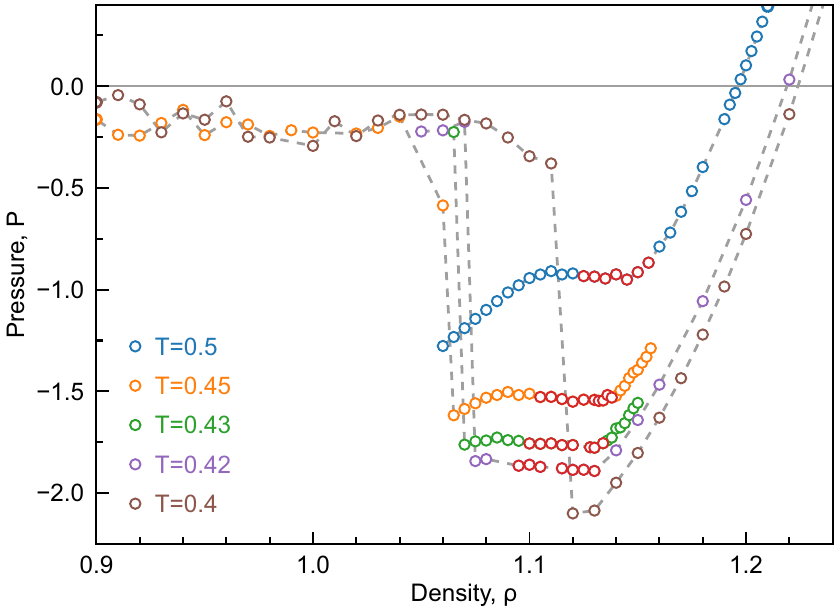}
    \caption{The isotherms of the LJ system at temperatures $T<T_{t2}$ in the high-density region.}
    \label{NMP-Figure9}
\end{figure}

This unexpected result is explained using the Gibbs rule: Due to
the reduced spatial dimension and since melting represents a
two-stage process, we obtain a couple of points where three phases
may exist simultaneously: hexatic-liquid-gas (at temperature
$T_{t1}$) and crystal-hexatic-gas (at lower temperature $T_{t2} <
T_{t1}$). This situation is caused by appearance of an additional
(hexatic) phase, inherent in 2D systems, and stands in contrast to
simple bulk systems, where only one triple point gas-liquid-solid
may exist. In the interval $T_t2<T_t1$, a continuous BKT
transition from the crystal to hexatic phase and a first-order
transition from hexatic to gas occur up to the crystal-gas
sublimation line, at which a first-order transition from crystal
to gas is observed on the isotherms. Simulation of first-order
gas-solid transitions at temperatures below $T_{t2}$ for finite
systems inevitably leads to the appearance of metastable states.
This is manifested by the Mayer-Wood loops, strongly extended due
to a large difference of densities on the isotherms. Moreover,
these loops turn out to have a complex structure, as given in
Fig.~\ref{NMP-Figure9}.

\begin{figure}[!t]
    \includegraphics[width=80mm]{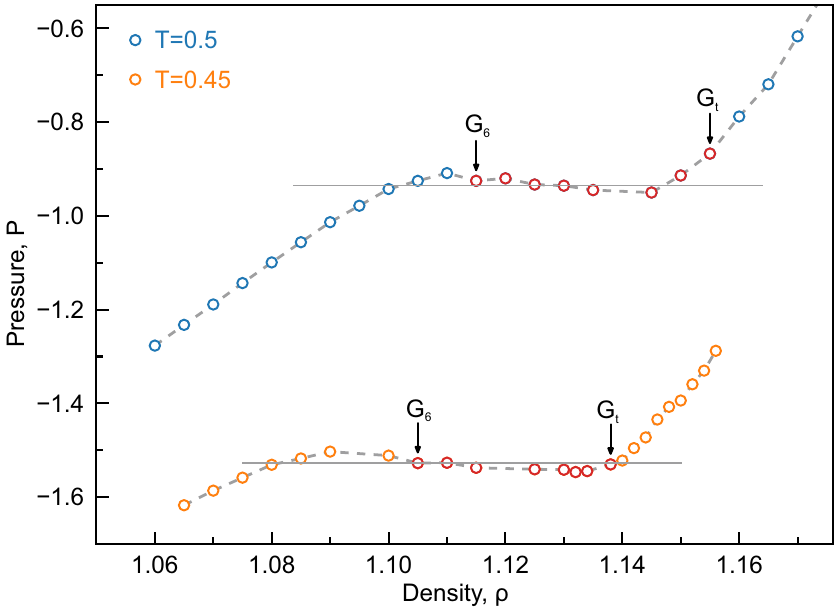}
    \caption{The change of the transition scenario from the third scenario (isotherm $T=0.5$) to a first-order transition (isotherm $T=0.45$) in the area of negative pressures at $T<T_{t2}$, which is related to suppression of the hexatic phase at $T=0.45$.}
    \label{NMP-Figure10}
\end{figure}

At high densities in the vicinity of the crystalline phase,
additional modulations of the loops may be interpreted as
two-stage melting of the metastable crystal appearing in the
equation of state. In this case, phase identification was
performed using analysis of the radial distribution functions,
behavior of the orientational and translational order parameters,
and their correlation functions. We observed a transition of the
metastable crystal to the hexatic phase through a continuous BKT
transition with a subsequent first-order transition from hexatic
to gas. With a decrease in temperature, the region of the hexatic
phase narrows on the phase diagram, and disappears completely at
$T=0.45$, which leads to one first-order crystal-gas transition,
as shown in Fig.~\ref{NMP-Figure10}.

\begin{figure}[!t]
    \includegraphics[width=80mm]{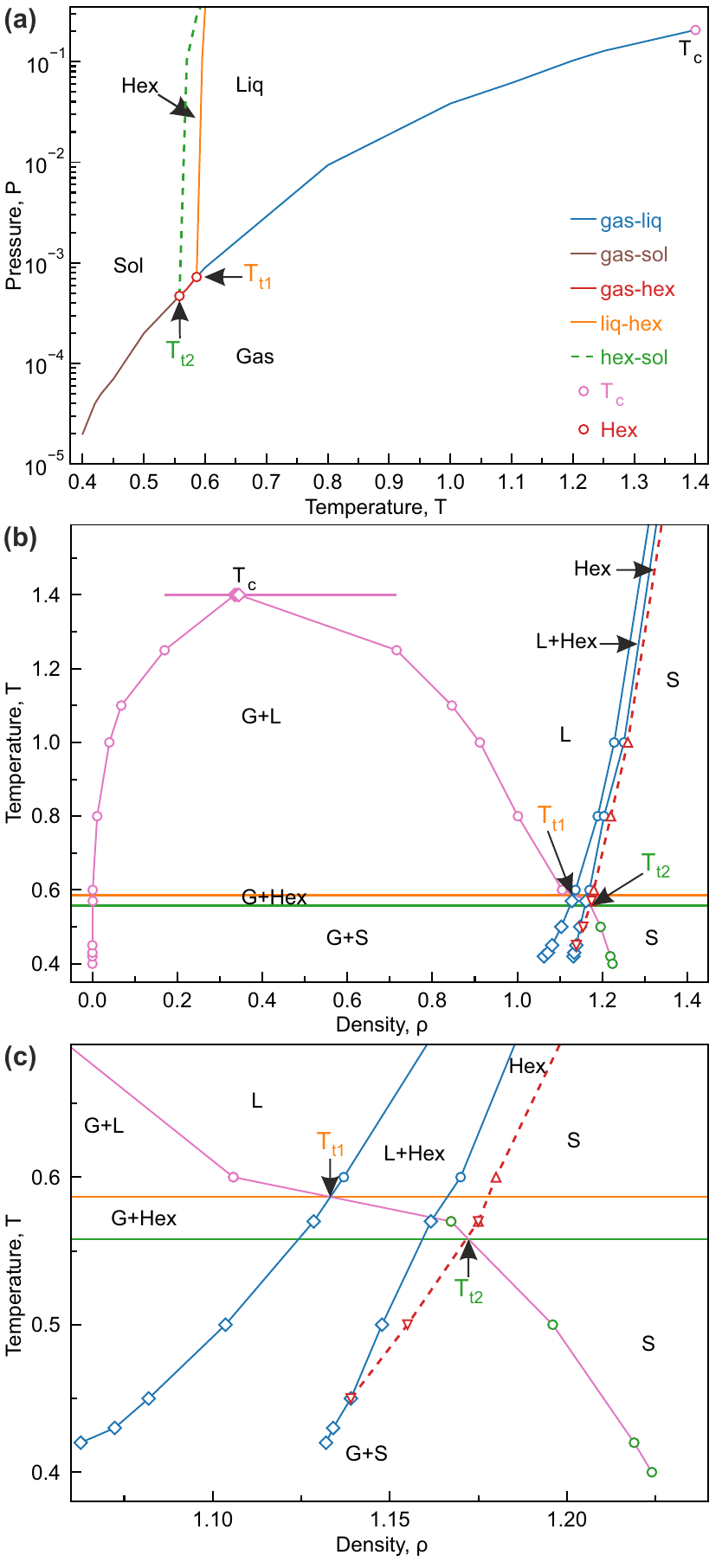}
    \caption{\textbf{The phase diagram of the LJ12-3 system:}
    (a) The phase diagram in the ($P-T$) coordinates.
    (b) The phase diagram in the ($\rho -T$) coordinates.
    $T_{CP}$ is a gas-liquid critical point,
    $T_{t1}$ is a hexatic-liquid-gas triple point, and $T_{t2}$ is a crystal-hexatic-gas triple point.
    Gas, liquid, hexatic phase and crystal are denoted as G, L, Hex, and S, respectively.
    The regions of their coexistence are marked as G+L, G+Hex, and G+S.
    The solid and dashed lines correspond to a first-order transition and to continuous BKT crystal-hexatic phase transition, respectively.
    The two-stage crystal-gas transition in the metastable area of the hexatic phase and crystal at $T<T_{t2}$ are denoted with open symbols.
    (c) The enlarged region of the phase diagram in the area of triple points $T_{t1}$ and $T_{t2}$.}
    \label{NMP-Figure11}
\end{figure}

For the case of isotropic dipolar attraction $m=3$, the phase
diagrams in different coordinates are presented in
Figs.~\ref{NMP-Figure11}(a,b). Here, the lines of phase
coexistence, critical, and triple points are shown in
Fig.~\ref{NMP-Figure11}(a). In the density-temperature plane, a
two-stage crystal-gas transition occurs in the metastable area of
the hexatic phase and crystal at $T<T_{t2}$, see Fig.
\ref{NMP-Figure11}(b). In Fig.~\ref{NMP-Figure11}(c) an
enlargement of the phase diagram is represented the behavior of
different domain boundaries on the phase diagram near triple
points $T_{t1}$ and $T_{t2}$. We note in conclusion that, with an
increase in the number of particles in the system, the loops
corresponding to gas-crystal transition should flatten. This leads
to the disappearance of the metastable area of the crystal and to
the first-order gas-crystal transition, without the modulation of
the equation of state mentioned above.

\section{Conclusions}

In this work, we studied the evolution of the phase diagrams and
melting scenarios of two dimensional systems of particles
interacting via generalized LJ potential with a different
attraction range, whereas the repulsion branch was fixed.
Gas-liquid transition was studied using analysis of the equation
of state and the phase identification method. The results obtained
by the two methods stand in good agreement indicating their
consistency. The drop in the attraction range reduces the
gas-liquid coexistence region and temperatures of the critical and
triple points.

Melting at high temperatures (and high densities) is found to
behave like a system of soft disks $1/r^{12}$, via the third
scenario. However, at low melting temperatures, a change in the
melting scenario was identified from the third scenario to a
first-order transition (without the hexatic phase) in the systems
with $m=6$, $9$, and $11$. The temperature of the change in the
scenarios is shifted toward lower temperatures with an increase in
the attraction range (corresponding to the decrease in $m$).
Analysis of the case $m=9$ (LJ12-9), very close to that studied in
Ref.~\cite{10.1103/physrevx.9.031032}, demonstrated that for
short-range attraction in complex solvents we should witness the
third melting scenario.

The largest region of gas-liquid coexistence was observed in the
phase diagram in the $\rho-T$ coordinates in the case of $m=3$.
Due to soft long-range isotropic dipolar attraction, we
unexpectedly saw two triple points at $T_{t1}$ and $T_{t2}$,
corresponding to hexatic-liquid-gas and to crystal-hexatic-gas
equilibrium, respectively. On the isotherms in the interval
between two triple points $T_{t2}$ and $T_{t1}$, we observed a
continuous BKT transition from crystal to hexatic phase and a
first-order transition from hexatic to gas. At the crystal-gas
sublimation line, a first-order transition from crystal to gas has
been identified.

The main conclusions related to the scenario of phase transitions
in a system with dipolar attraction can be tested in future
experiments with colloidal systems, wherein the tunable long-range
dipolar attraction can be created with rotating magnetic or
electric fields. A nontrivial question here is related to the role
of three-body forces, inherent in atomic materials and tunable
colloids \cite{10.1063/1.5131255, 10.1016/j.jcis.2021.09.116}.
LJ-like interactions can be created with magic hodographs of
rotating electric or magnetic fields \cite{10.1039/d0sm01046d}.
However, three-body interactions in this case behave as
$\propto1/r^6$, with the same asymptotics as pairwise potential.
Due to this, the scenario of melting can change, and we leave the
corresponding study to future work.


\acknowledgments
The study was supported by the Russian Science Foundation, Grant No. 19-12-00092 (MD simulations to obtain phase diagrams) and 20-12-00356 (analysis with phase identification method).
This work was carried out using computing resources of the federal collective usage center ``Complex for simulation and data processing for mega-science facilities'' at NRC ``Kurchatov Institute'',
http://ckp.nrcki.ru, and supercomputers at Joint Supercomputer Center of the Russian Academy of Sciences (JSCC RAS).

E.N.T., Y.D.F., and E.A.G. performed MD simulations to obtain phase diagrams,
N.A.D., P.A.L., and N.P.K. obtained condensate-gas binodals with phase identification method,
E.N.T., V.N.R.,  E.E.T., N.P.K., and S.O.Y. analysed and discussed the results,
E.N.T., V.N.R., and S.O.Y. wrote the manuscript,
V.N.R. and S.O.Y. conceived and directed the study.

\bibliography{Ref-NMP}

\end{document}